\documentclass[draft,grl]{agu2001}
\usepackage{graphicx}
\usepackage{lineno}
\linenumbers

\authorrunninghead{YERMOLAEV, LODKINA, NIKOLAEVA AND YERMOLAEV}

\titlerunninghead{FLOOR IN IMF}

\authoraddr{Yu. I. Yermolaev
Space Plasma Physics Department, Space Research Institute, 
Russian Academy of Sciences, Profsoyuznaya 84/32, Moscow 117997, Russia. 
(yermol@iki.rssi.ru)} 
\authoraddr{I. G. Lodkina
Space Plasma Physics Department, Space Research Institute, 
Russian Academy of Sciences, Profsoyuznaya 84/32, Moscow 117997, Russia.}
\authoraddr{N. S. Nikolaeva
Space Plasma Physics Department, Space Research Institute, 
Russian Academy of Sciences, Profsoyuznaya 84/32, Moscow 117997, Russia.}
\authoraddr{M.Yu. Yermolaev
Space Plasma Physics Department, Space Research Institute, 
Russian Academy of Sciences, Profsoyuznaya 84/32, Moscow 117997, Russia.}

\begin{document}

%
%
%
%
%

%
%

\title{The "floor" in the interplanetary magnetic field: Estimation on the basis of 
       relative duration of ICME observations in solar wind during 1976-2000
}

%
%



\author{Yu. I. Yermolaev, I.G. Lodkina, N.S. Nikolaeva, and M. Yu. Yermolaev} 
\affil{Space Plasma Physics Department, Space Research Institute, 
Russian Academy of Sciences, Profsoyuznaya 84/32, Moscow 117997, Russia}

\begin{abstract}
To measure the floor in interplanetary magnetic field and estimate the time-invariant open magnetic flux of Sun, it is necessary to know a part of magnetic field of Sun carried away by CMEs. In contrast with previous papers, we did not use global solar parameters: we identified different large-scale types of solar wind for 1976-2000 interval, obtained a fraction of interplanetary CMEs (ICMEs) and calculated magnitude of interplanetary magnetic field $B$ averaged over 2 Carrington rotations. The floor of magnetic field is estimated as $B$ value at solar cycle minimum when the ICMEs were not observed and it was calculated to be $4.65\pm6.0$ nT. Obtained value is in a good agreement with previous results. 
\end{abstract}
%
%

%

\begin{article}

%
%

\section{Introduction}

One of the basic problems of physics of the Sun and heliosphere is the estimation of magnetic flux of the Sun which is carried away by the solar wind. Several models (see, for instant, papers by
\cite{McComasetal1992,WebbHoward1994,OwensCrooker2006,OwensCrooker2007,Owensetal2008}
and references therin) suggest that there is a minimum magnetic field, floor of the open magnetic flux, which is a constant (time-independent) value and may be estimated on the basis of measurements during  solar minimum because during solar maximum the coronal mass ejections (CME) carry away an additional (time-dependent) part of magnetic field, closed solar magnetic flux. Thus, if the "floor" really exists, it can be found out and estimated experimentally. 

Estimations of the floor of magnetic field have been obtained by 2 methods.

(1) Magnitude $B$ of interplanetary magnetic field (IMF) is compared with phase of solar cycle (sunspot number) and  this dependence is extrapolated in conditions when sunspot number is minimal. It is supposed that the value of a magnetic field obtained by this way is a value of "floor". This method allows authors to obtain the estimations $\sim5$ nT during 40 years 
\citep{Richardsonetal2002}
and $\sim 4.6$ nT during last 130 years 
\citep{SvalgaardCliver2007}.

(2) Magnitude $B$ of IMF is compared with daily CME rate observed with LASCO coronagraph on SOHO spacecraft 
\citep{OwensCrooker2006}
and this dependence is extrapolated to values when CME rate is zero. The floor value $4.0\pm0.3$ nT has been obtained by second methods 
\citep{Owensetal2008}.

Both methods use global characteristics of the Sun for "floor" estimation, and IMF measurements near the Earth reflects only conditions near the ecliptic plane or field in the streams generating by the low-latitude regions of the Sun. If there is a latitude anisotropy of characteristics of the Sun (sunspot number and ÑÌÅ distribution on the solar disk) it leads to an error of floor estimations. Therefore ÑÌÅ contribution to a magnetic flux near the Earth is better for estimating on the basis of the characteristics of a solar wind measured near the Earth, instead of on the Sun. 

In this paper we suggest to use a fraction of interplanetary CMEs (ICME) in observation of the solar wind and IMF measured near the Earth (OMNI database) in 1976-2000 as a measure of contribution of CMEs in outflow of solar magnetic flux. 

\section{Method of data processing}

On the basis of the OMNI database of interplanetary measurements near the Earth (http://omniweb.gsfc.nasa.gov) 
\citep{KingPapitashvili2004}, 
we identified solar wind (SW) types for all time intervals during 1976 – 2000. Our classification includes quasi-steady types: (1) heliospheric current sheet (HCS), (2) slow and (3) fast SW streams, and disturbed types: (4) corotating interaction regions (CIR), (5) sheath and (6) magnetic cloud (MC) and (7) ejecta (Ejecta) as well as (8) direct and (9) reverse interplanetary shocks (see papers by 
\cite{Yermolaevetal2009a,Yermolaevetal2009b} 
and catalogue on site ftp://ftp.iki.rssi.ru/pub/omni/). Our analysis showed that MC was observed during $\sim2\%$ of total time of observations, Ejecta $\sim20\%$, Sheath before MC and Ejecta $\sim9\%$, CIR $\sim10\%$, HCS $\sim6\%$, Fast SW streams $\sim21\%$ and Slow ones $\sim31\%$. Identification of MC and Ejecta events was carried out according to the standard criteria. MCs are a subclass of Ejecta and in contract to Ejecta, MCs have higher and more regular magnetic field, thus observant distinctions can be connected, as with intensity of CME on the Sun, and as with a trajectory of a spacecraft relative to an axis of a magnetic flux rope in MC/EJECTA 
\citep{Burlaga1991,RichardsonCane1995,RussellMulligan2002,CaneRichardson2003}. 
Thus, on the average the solar wind during 1976 - 2000 consisted of $\sim22\%$ of ICMEs which include both MCs and Ejecta. 

Fig.1 shows temporal distribution of Sheath+MC (left panel) and Sheath+Ejecta (right panel) observations (Carrington Rotations, CR nos. 1636-1971). It is important to note that data gap of OMNI database (white regions in Fig.1) is about $48\%$ of total time interval 1976-2000. Several 
CRs included less 100 hours of observation ($<15\%$ of CR duration) and these CRs have been excluded from our analysis. We calculated average magnitude $B$ of IMF and ICME fraction $F$ (ratio of ICME duration to total duration of interplanetary field and plasma data). Time of averaging was duration of 2 CRs to increase statistics. These parameters $B$ and $F$ are analyzed in next Section of the paper. 

 \begin{figure}
   \centering
   \includegraphics[width=0.8\textwidth]{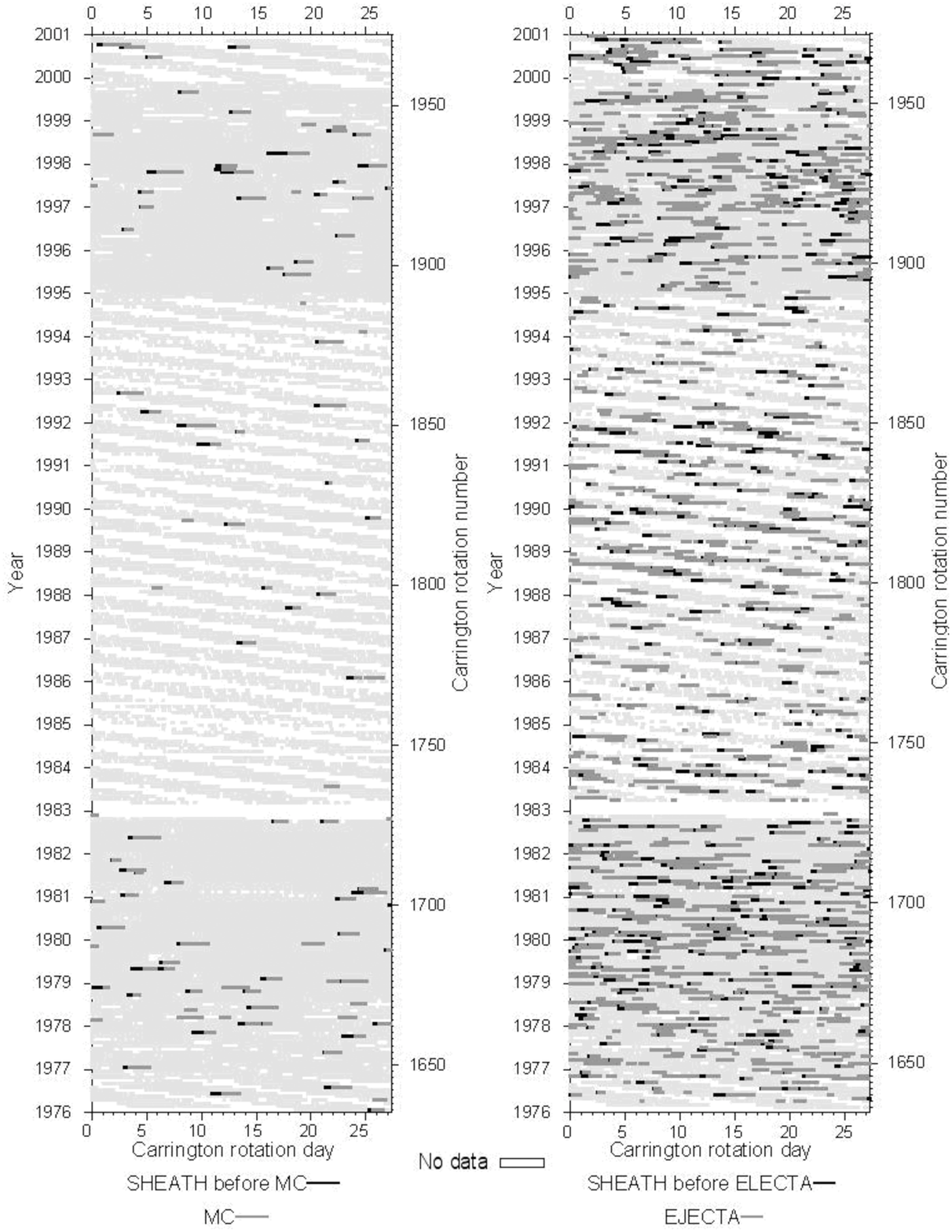}
      \caption{Temporal distribution of MC+Sheath (left panel) and Ejecta+Sheath (right panel) observations
               during 1976-2000      
       \citep{Yermolaevetal2009b}.
       }
         \label{Fig1}
   \end{figure}

\section{Results}

Dependence of average magnitude $B$ of IMF on ICME fraction $F$ is presented in Fig.2. Upper panel shows 2 CR means of parameters for different time intervals: solar cycle minimum – open circles 1976-1977, open square 1984-1987 and open triangles 1994-1998 and solar cycle maximum - closed circles 1978-1983, closed square 1988-1993 and closed triangles 1999-2000. There is a significant difference between minimal and maximal phases of solar cycle and there is no difference between 21 and 22 cycles for the same phases. 

Bottom panel shows average values and dispersions of $B$ calculating in $F$ bins: 0.0- 0.05; 0.05-0.1 etc. Data in 0.0-0.15 interval have been approximated by linear functions ($B = 5.92 + 8.08\times F$ for maximum of solar cycle and $ B = 4.65 + 6.89\times F$  for minimum, respectively), and in 0.2-0.45 interval – by constant values ($B = 6.81$ and $B = 5.50$ nT, respectively). For solar cycle minimum $B = 4.65 \pm 0.6$ nT when $F = 0$ and this value is our estimation of the floor in the interplanetary magnetic field during 1976-2000.  

\begin{figure}
   \centering
   \includegraphics[width=0.7\textwidth]{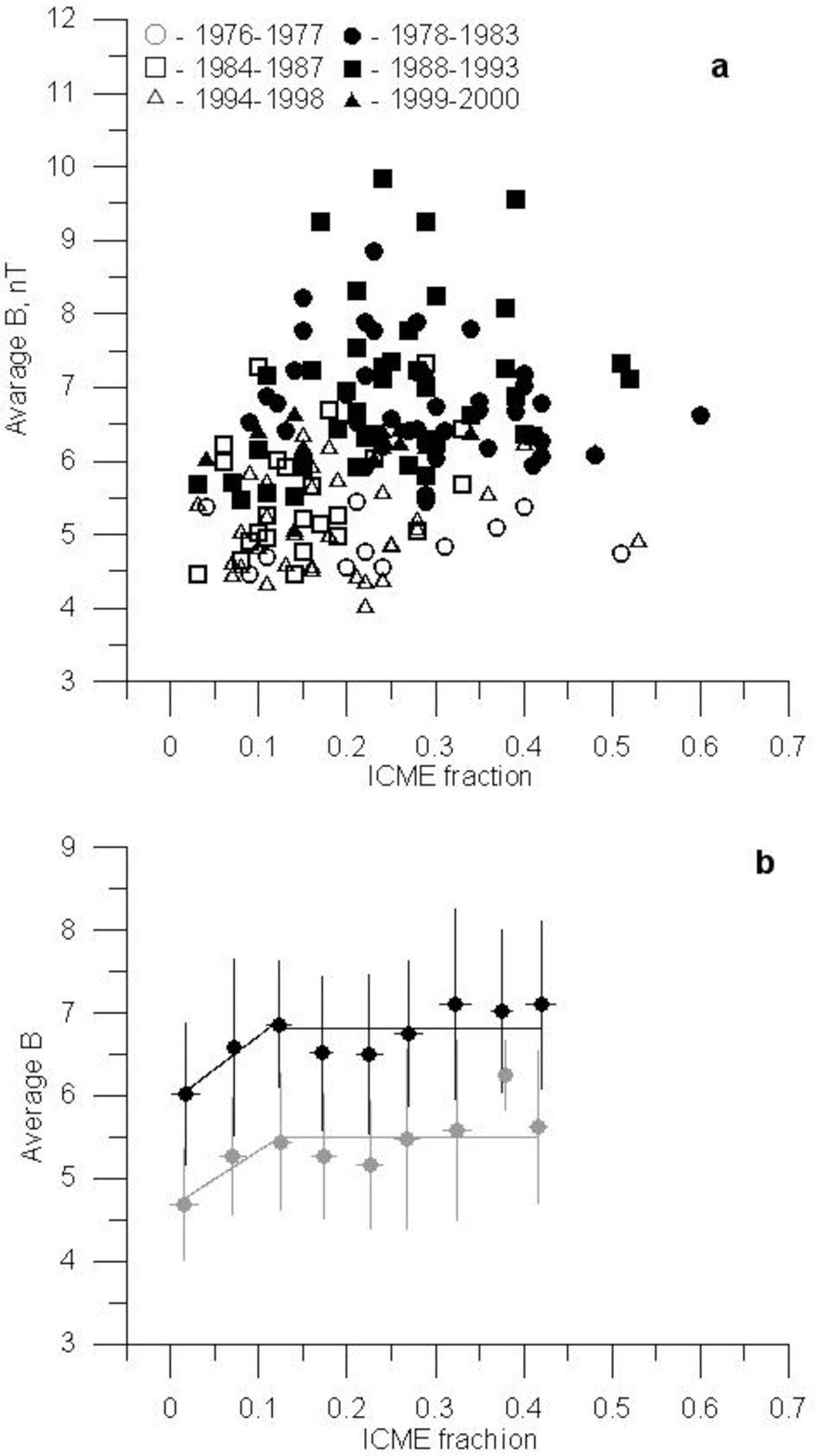}
      \caption{Dependence of average magnetic field $B$ on ICME fraction: (a) 2 Carrington rotation average for different phases of solar cycle, (b) avarave over ICME fraction bins for maximum (black line) and minimum (grey line) of solar cycle 
       }
         \label{Fig2}
   \end{figure}

\section{Discussion and conclusions} 
Using dependence of the magnitude of interplanetary magnetic field $B$ averaged over 2 Carrington rotations on the ICME fraction in measurements of solar wind $F$ during 1976-2000, we made estimation of the floor in IMF and found  that $B = 4.65 \pm 0.6$ nT at minimum of solar activity and at $F = 0$. We did not find a significant difference between the floors during minima of different solar cycles. Our results with spread are in a good agreement with previous results: $\sim5$ nT 
\citep{Richardsonetal2002}, 
$\sim 4.6$ nT 
\citep{SvalgaardCliver2007} 
and $4.0 \sim 0.3$ nT 
\citep{Owensetal2008}. 
Our result was obtained for 21 and 22 solar cycles and does not contradict to result $4.0 \sim 0.3$ nT 
\citep{Owensetal2008} 
obtained for the end of 23 solar cycle and hypothesis that that floor in IMF may decrease in 23 cycle 
\citep{McCracken2007,Owensetal2008}.

\begin{acknowledgements} 
       We are grateful to OMNI team for possibility to use the database for our investigations. Work was in part supported by RFBR, grant 07-02-00042.  
\end{acknowledgements}

\end{article}

\end{document}